\documentclass[12pt]{article}
\usepackage{graphicx}

\begin{document}

\newcommand{\bm}[1]{\mbox{\boldmath$#1$}}

\begin{center}
{\Large\bf On exciton decoherence in quantum dots}
\end{center}

\begin{center}
L.Jacak, A.Janutka, P.Machnikowski, A.Radosz\\
{\small\em Institute of Physics, Wroc{\l}aw University of Technology,
Wybrze\.ze Wyspia\'nskiego 27, 50-370 Wroc{\l}aw, Poland}\\ 
J.Krasnyj\\
{\small\em Institute of Mathematics, University of Opole,
Oleska 48, 45-051 Opole, Poland} 
\end{center}

\noindent
{\bf Abstract:}
The effects resulting due to dressing of an exciton with phonons
are analysed as the source of unavoidable decoherence of orbital 
degrees of freedom in quantum dots. The dressing with longitudinal optical
phonons results in energetic shift of order of a few meV even of
the ground state of exciton in a state-of-the-art 
InAs/GaAs dot and the mediating role 
of longitudinal acoustical phonons is essential in this process.  
The characteristic time needed for dressing of the exciton with optical 
phonons is of a picosecond order. That time can be regarded as the
lower limit for decoherence for optically driven quantum gates
employing self--assembled quantum dot structures.\\


\noindent
{\bf e-mail:} radosza@if.pwr.wroc.pl,\\
\noindent
{\bf tel:} tel. (+48)71 3202361

\newpage

\section{Introduction}

We consider an exciton, i.e. an electron--hole pair created in a quantum 
 dot (QD) by extremely short optical pulse (Borri 2001, Kikkawa 1988).
 The rapidly created bare exciton is further dressing 
 with phonons, which results in arising a composite quasiparticle being 
 the coherent mixture of the electron--hole pair and phonons. Its energy
 is lower than the bare exciton energy. In the case of interaction with 
 only optical phonons this composite particle can be called 
 exciton--polaron in analogy to the Frolich electron--polaron. 
 In a realistic model much weaker deformation effects responsible for 
 the interaction with acoustical phonons are included. 

The observation of the exciton dressing with phonons is possible when
 the dipole coupling with the electromagnetic
 field is sufficiently strong to preserve a non adiabatic (i.e. rapid)
 exciton creation (cf. 0.2ps laser pulse in Ref. (Borri 2001)) 
 much faster than dressing with phonons (of ps scale as we will show below).   
 We consider a weakly polar medium for QDs (GaAs) in which the interaction
 of electrons with longitudinal optical (LO) phonons is dominating 
 in terms of the influence on the quasiparticle energy shift. Much
 weaker interaction with gap--less longitudinal acoustic (LA) phonons plays 
 an important role in kinematics of the dressing process,
 being a channel of energy transfer from exciton. 
 In order to determine the limit of decoherence time of the optical
 excitations in QDs, this LA channel for dressing exciton with LO/LA phonons
 must be included.
 
We describe and analyse the dressing and the relaxation of exciton
 problems in detail for typical 
 self--assembled strain induced InAs/GaAs QDs (Jacak 1998) within the
 Green function formulation. We model the QD by the parabolic confinement 
 potential.                     

\section{Phonon dressing description}

In order to investigate the time evolution of the lowest exciton state,
 we consider the Hamiltonian describing a single electron and a hole
 interacting with phonons
\begin{eqnarray}
{\cal H}={\cal H}_{e}({\bf r}_{e})+{\cal H}_{h}({\bf r}_{h})
-\frac{e^{2}}{\epsilon_{0}|{\bf r}_{e}-{\bf r}_{h}|}
+\sum_{{\bf k}s}\hbar\omega_{{\bf k}s}b_{{\bf k}s}^{\dagger}b_{{\bf k}s}
-\frac{e}{N^{1/2}}
\left(\frac{2\pi\hbar\Omega}{\upsilon\tilde{\epsilon}}\right)^{1/2}
\sum_{\bf k}\frac{1}{k}
\nonumber\\
\times
(b_{{\bf k}o}+b_{-{\bf k}o}^{\dagger})
({\rm e}^{{\rm i}{\bf kr}_{e}}-{\rm e}^{{\rm i}{\bf kr}_{h}})
-\frac{1}{N^{1/2}}
\left(\frac{\hbar}{2Mc_{a}}\right)^{1/2}
\sum_{\bf k}k^{1/2}(b_{{\bf k}a}+b_{-{\bf k}a}^{\dagger})
(\sigma_{e}{\rm e}^{{\rm i}{\bf kr}_{e}}
-\sigma_{h}{\rm e}^{{\rm i}{\bf kr}_{h}}).
\label{Hamil}
\end{eqnarray}
Here $b_{{\bf k}s}^{\dagger}(b_{{\bf k}s})$ denote the phonon creation 
 (annihilation) operators, ($s=o$ for LO, $s=a$ for LA), $\Omega$ denotes 
 the frequency of the zero wave vector LO phonons, 
 $c_{a}$ -- the LA phonon frequency, $\sigma_{e,h}$ -- deformation constants
 for the electron and hole, $M$ -- the mass of ions in the elementary cell, 
 $\upsilon$ -- the elementary cell volume, $N$ -- the number of cells
 in crystal, $\tilde{\epsilon}=(1/\epsilon_{\infty}-1/\epsilon_{0})^{-1}$ 
 -- the effective dielectric constant. The electron (hole) part of
 the Hamiltonian 
\begin{eqnarray}
{\cal H}_{i}({\bf r}_{i})=-\frac{\hbar^{2}}{2m_{i}^{*}}
\nabla_{i}^{2}
+\frac{1}{2}m_{i}^{*}(\omega_{0}^{i})^{2}
r_{\bot i}^{2}
+\frac{1}{2}m_{i}^{*}(\omega_{z}^{i})^{2}
z_{i}^{2}
\end{eqnarray}
(where $i=e,h$, $r_{\bot i}^{2}=x_{i}^{2}+y_{i}^{2}$) contains 
 the vertical and horizontal confinement potentials. There are characteristic
 confinement sizes $l_{i}=\sqrt{\frac{\hbar}{m_{i}^{*}\omega_{0}^{i}}}$,
 $l_{iz}=\sqrt{\frac{\hbar}{m_{i}^{*}\omega_{z}^{i}}}$ 
 ($\omega_{0}^{i}\ll\omega_{z}^{i}$). Finding the eigenfunctions 
 $\Phi_{n}({\bf r}_{e},{\bf r}_{h})$ of the exciton Hamiltonian 
\begin{equation}
{\cal H}_{ex}={\cal H}_{e}({\bf r}_{e})+{\cal H}_{h}({\bf r}_{h})
-\frac{e^{2}}{\epsilon_{0}|{\bf r}_{e}-{\bf r}_{h}|},
\end{equation}
we write the Hamiltonian (\ref{Hamil}) in the second quantisation
 representation in the form of the sum of unperturbed and interaction
 parts
\begin{eqnarray}
{\cal H}=\sum_{n}\hbar E_{n}a_{n}^{\dagger}a_{n}
+\sum_{{\bf k},s}\hbar\omega_{{\bf k}s}b_{{\bf k}s}^{\dagger}b_{{\bf k}s}
+\frac{1}{N^{1/2}}
\sum_{n_{1},n_{2},{\bf k},s}F_{s}(n_{1},n_{2},{\bf k})
a_{n_{1}}^{\dagger}a_{n_{2}}
(b_{{\bf k}o}+b_{-{\bf k}o}^{\dagger}),
\label{Hamil1}
\end{eqnarray}
where
\begin{eqnarray}
F_{o}(n_{1},n_{2},{\bf k})=
-\frac{e}{k}
\left(\frac{2\pi\hbar\Omega}{\upsilon\tilde{\epsilon}}\right)^{1/2}
\int\Phi_{n_{1}}^{*}({\bf r}_{e},{\bf r}_{h})
({\rm e}^{{\rm i}{\bf kr}_{e}}-{\rm e}^{{\rm i}{\bf kr}_{h}})
\Phi_{n_{2}}({\bf r}_{e},{\bf r}_{h}){\rm d}^{3}{\bf r}_{e}
{\rm d}^{3}{\bf r}_{h},
\nonumber\\
F_{a}(n_{1},n_{2},{\bf k})=
-\left(\frac{\hbar k}{2Mc_{a}}\right)^{1/2}
\int\Phi_{n_{1}}^{*}({\bf r}_{e},{\bf r}_{h})
(\sigma_{e}{\rm e}^{{\rm i}{\bf kr}_{e}}
-\sigma_{h}{\rm e}^{{\rm i}{\bf kr}_{h}})
\Phi_{n_{2}}({\bf r}_{e},{\bf r}_{h}){\rm d}^{3}{\bf r}_{e}
{\rm d}^{3}{\bf r}_{h}.
\label{r0}
\end{eqnarray}

The linear response of the exciton to the electromagnetic field is described
 by using the retarded Green function of the electric current operators
\begin{eqnarray}
{\bf j}(t)
\sim\sum_{n}
{\bf d}_{n}[a_{n}(t)+a_{n}^{\dagger}(t)],
\end{eqnarray}
here ${\bf d}_{n}$ denotes an effective dipole moment of the exciton
 (Elliott 1957, Mahan 2000).
 The spectral density of the linear response function 
 is determined by the causal one--particle functions
 $G_{nn^{'}}(t)=-{\rm i}\langle T\{a_{n}(t)a_{n^{'}}^{\dagger}\}\rangle$
 (where the average is taken with Hamiltonian (\ref{Hamil1}))
 which we evaluate solving the Dyson equation. We calculate
 the spectral function $A(\omega)\equiv-2{\rm Im}G^{r}_{00}(\omega)=
-2{\rm Im}G_{00}(\omega+{\rm i}0^{+})$ and its 
 time course $A(t)=(2\pi)^{-1}\int_{-\infty}^{\infty}A(\omega)
{\rm e}^{-{\rm i}\omega t}{\rm d}\omega$.
 In order to determine the one--particle causal function, we evaluate the 
 components of the mass operator using standard method of the solution of
 the Green function equations of motion
\begin{eqnarray}
\left({\rm i}\hbar\frac{{\rm d}}{{\rm d}t}-E_{n}\right)G_{nn^{'}}(t)
-\frac{1}{N^{1/2}}\sum_{n_{1},{\bf k}s}F_{s}(n,n_{1},{\bf k})
R_{n_{1}n^{'}}({\bf k}s,t)=\delta(t)\delta_{nn^{'}},
\label{rA1}
\end{eqnarray}
where
\begin{equation}
R_{n_{1}n^{'}}({\bf k}s,t)=-\frac{{\rm i}}{\hbar}\langle T\{a_{n_{1}}(t)
[b_{{\bf k}s}(t)+b_{-{\bf k}s}^{\dagger}(t)]
a_{n^{'}}^{\dagger}(0)\}\rangle,
\end{equation}
with additional condition $\langle\tilde{\varphi}_{{\bf k}s}(t)\rangle
\equiv\langle b_{{\bf k}s}(t)
+b_{-{\bf k}s}^{\dagger}(t)\rangle=0$ (Martin 1959, Engelsberg 1963).
 Its Fourier transform takes the form of the Dyson equation 
\begin{equation}
G_{nn^{'}}(\omega)=G_{nn_{1}}^{(0)}(\omega)
+\sum_{n_{1}n_{2}}G_{nn_{1}}^{(0)}(\omega)
\Sigma_{n_{1}n_{2}}(\omega)G_{n_{2}n^{'}}(\omega).
\label{Dyson}
\end{equation} 
Neglecting the off diagonal Green functions, one finds
 the diagonal elements of the mass operator 
\begin{eqnarray}
\Sigma_{nn}(\omega)=
\frac{1}{N}
\sum_{{\bf k}s}
\sum_{n_{1}}
F_{s}(n,n_{1},{\bf k})F_{s}(n_{1},n,-{\bf k})
\left\{\frac{
\Gamma_{n_{1}n,{\bf k}s}(\omega-\omega_{{\bf k}s},-\omega_{{\bf k}s})
[1+n(\hbar\omega_{{\bf k}s})]}{\hbar(\omega-\omega_{{\bf k}s})-E_{n_{1}}
-\Sigma_{n_{1}n_{1}}(\omega-\omega_{{\bf k}s})}
\right.\nonumber\\\left.
+\frac{
\Gamma_{n_{1}n,{\bf k}s}(\omega+\omega_{{\bf k}s},\omega_{{\bf k}s})
n(\hbar\omega_{{\bf k}s})}{\hbar(\omega+\omega_{{\bf k}s})-E_{n_{1}}
-\Sigma_{n_{1}n_{1}}(\omega+\omega_{{\bf k}s})}
\right\}.
\label{rA5}
\end{eqnarray}
Here $n(x)=({\rm e}^{x/T}-1)^{-1}$, and
 $\Gamma_{nn^{'},{\bf k}s}(\omega_{1},\omega_{2})$
 denotes the Fourier transform of a vertex function 
\begin{eqnarray}
\Gamma_{nn^{'},{\bf k}s}(t,t^{'})
\equiv-\frac{\delta G_{nn^{'}}^{-1}(t)}{
F_{s}(n,n^{'},{\bf k})
\delta\tilde{\varphi}_{{\bf k}s}(t^{'})}.
\end{eqnarray}
satisfying $\Gamma_{nn,{\bf k}s}(\omega,\omega_{1})\approx 
1-\hbar^{-1}{\rm d}\Sigma_{nn}/{\rm d}\omega$ when consider
 terms of the lowest order in ${\bf k}$. Let us denote 
 real end imaginary part of the mass operator, respectively,
 ${\rm Re}\Sigma_{00}(\omega)\equiv\Delta(\omega)$,  
 ${\rm Im}\Sigma_{00}(\omega)\equiv-\gamma(\omega)$. 

We may solve the mass operator equation (\ref{rA5})
 with the assumption that the dressed exciton is 
 a durable composite particle which is undamped (Suna 1964). 
 It corresponds to the strictly solvable one level model 
 of (Krummheuer 2002). We expect in this case
 that the exciton Green function has a real pole at frequency corresponding 
 to the exciton--polaron energy. In the frequency region in vicinity 
 of this pole, the imaginary part of the mass operator can be neglected
 compared to the real part, which allows us to write the self--consistent
 equation for $\Delta(\omega)$ (assuming $\Delta(\omega)\simeq-\Delta
={\rm const}$)
\begin{eqnarray}
-\Delta\simeq
\frac{\hbar}{N}\sum_{\bf k}J_{00}^{o}({\bf k})
\left\{\frac{[1+n(\hbar\omega_{{\bf k}o})]}
{E_{0}-\tilde{E}_{n}-\hbar\omega_{{\bf k}o}}
+\frac{n(\hbar\omega_{{\bf k}o})}
{E_{0}-\tilde{E}_{n}+\hbar\omega_{{\bf k}o}}\right\}
\label{selfcon}
\end{eqnarray}
which may be solved exactly
($\tilde{E}_{n}=E_{0}-\Delta$, 
 $J_{nn^{'}}^{s}({\bf k})=F_{s}(n,n^{'},{\bf k})F_{s}(n^{'},n,-{\bf k})$).
 The energy shift due to the interaction with
 LA phonons is negligible. One can calculate the form factors
\begin{eqnarray}
J_{00}^{o}({\bf k})=\frac{\pi e^{2}\hbar\Omega}{
18\upsilon\tilde{\epsilon}}(L_{e}^{2}-L_{h}^{2})^{2}
{\rm e}^{-l_{\bot}k_{\bot}^{2}/2-l_{z}k_{z}^{2}/2},
\nonumber\\
J_{00}^{a}({\bf k})=\frac{\hbar k}{2Mc_{a}}(\sigma_{e}-\sigma_{h})^{2}
{\rm e}^{-l_{\bot}k_{\bot}^{2}/2-l_{z}k_{z}^{2}/2},
\end{eqnarray}
(here $l=l_{e}\approx l_{h}$, $l_{z}=l_{ze}\approx l_{zh}$, 
 $L_{e}=\sqrt{\frac{\hbar}{m_{e}^{*}\omega_{m}}}$,
 $L_{h}=\sqrt{\frac{\hbar}{m_{h}^{*}\omega_{m}}}$, and 
 $\omega_{m}=\sqrt{\frac{m_{e}^{*}}{m}(\omega_{0}^{e})^{2}+
\frac{m_{h}^{*}}{m}(\omega_{0}^{h})^{2}}$, 
$m=m_{e}^{*}+m_{h}^{*}$) (Davydov 1976). 

However, we intend to describe the dressed 
 exciton as a damped quasiparticle at non zero temperature range.
 When include many exciton energy levels,
 the damping is a consequence of the inclusion of phonon dispersion which 
 is responsible for the interaction between the dressed exciton and the phonon
 subsystem (Davydov 1972). We solve (\ref{rA5}) in two iteration steps 
 in the vicinity of $\hbar\omega=E_{0}$, taking in the first approximation
 the vertex functions $\Gamma_{n_{1}n^{'},{\bf k}s}(\omega,\omega_{1})=1$,
 as done in (Moskalenko 1968). Starting from
 $\Delta_{(0)}(\omega)=0$, $\gamma_{(0)}(\omega)=0^{+}$ 
 and assuming that $\Delta(\omega)\simeq{\rm const}$,
 we find in the first iteration step
\begin{eqnarray}
\Delta_{(1)}(\omega)\simeq\Delta_{(1)}(E_{0}/\hbar)=-\Delta^{'}=
\frac{\hbar}{N}\sum_{\bf k}\frac{b_{n}J_{0n}^{o}({\bf k})}
{E_{0}-E_{n}-\hbar\omega_{{\bf k}o}},
\nonumber\\
\gamma_{(1)}(\omega)=\frac{\hbar\pi}{N}\sum_{\bf k}J_{00}^{a}({\bf k})
\left\{[1+n(\hbar\omega_{{\bf k}a})]
\delta(\hbar\omega-E_{0}-\hbar\omega_{{\bf k}a})
+n(\hbar\omega_{{\bf k}a})
\delta(\hbar\omega-E_{0}+\hbar\omega_{{\bf k}a})\right\}.
\end{eqnarray}
Here $b_{n}$ denotes the $n$--th level degeneracy rank ($b_{n}=1,2$).
 The optical phonon contribution to $\gamma(\omega)$,
 is significant for $\hbar\omega\simeq E_{n}\pm\hbar\Omega$ only.
 We include the LO phonon influence on $\gamma(\omega)$, introducing
 an effective constant to $J_{00}^{a}({\bf k})$, which does not 
 influence the time of fast dressing with phonons but it determines 
 the relaxation time at non zero $T$. This constant will be estimated
 from the sum rule $|A(t=0)|=1$ for $T=0$. In the second step, the relation
 $\Delta_{(1)}(\omega)\gg\gamma_{(1)}(\omega)$ enables one 
 to neglect the $\gamma_{(1)}(\omega)$ in the denominators of (\ref{rA5}).
 It leads to 
\begin{eqnarray}
\Delta_{(2)}(\omega)\simeq\Delta_{(2)}(E_{0}/\hbar)=-\Delta^{''}\simeq
\Delta_{(2)}(\tilde{E}_{0}/\hbar)+\hbar^{-1}
\frac{{\rm d}\Delta_{(2)}}{{\rm d}\omega}|_{\omega=\tilde{E}_{0}/\hbar}
\Delta^{'}
=-\Delta^{'}Z^{-1},	  
\nonumber\\
\gamma_{(2)}(\omega)=\gamma_{(1)}(\omega+\Delta^{'}/\hbar),
\end{eqnarray}
where $\tilde{E}_{n}=E_{n}-\Delta^{'}$, 
 $\Delta_{(2)}(\tilde{E}_{0}/\hbar)=
 \frac{\hbar}{N}\sum_{\bf k}\frac{b_{n}J_{0n}^{o}({\bf k})}
 {\tilde{E}_{0}-\tilde{E}_{n}-\hbar\omega_{{\bf k}o}}=-\Delta^{'}$.
 Continuing the iteration process we would arrive to a self consistent 
 equation for $\Delta$ similar to (\ref{selfcon}), however, after
 infinite number of the approximations. We find 
\begin{eqnarray}
\gamma_{(2)}(x)
=Z^{-1}\alpha x^{3}{\rm e}^{-\beta x^{2}}f(x)\left\{-n(-x)\theta(-x)
+\left[1+n(x)\right]\theta(x)\right\}
\label{gam}
\end{eqnarray}
where $x=\hbar\omega-E_{0}+\Delta^{'}$, 
 $f(x)=\sum_{n=0}^{\infty}\frac{\left[
-\beta x^{2}(l_{z}^{2}/l_{\bot}^{2}-1)\right]^{n}}{(2n+1)n!}$,
 $Z$ denotes 
 the vertex function renormalisation factor
 ($Z=|1-\hbar^{-1}{\rm d}\Delta/{\rm d}\omega|^{-1}$).
 We write the spectral function in the form 
\begin{eqnarray}
A(\omega)=
Z\frac{\hbar 2\tau^{-1}(\omega)}{
[\hbar\omega-E_{0}+Z^{-1}\Delta^{'}]^{2}
+\tau^{-2}(\omega)},
\end{eqnarray}
where $\tau^{-1}(\omega)=Z\gamma(\omega)$ (Langreth 1964, Mahan 1966). Since 
 at zero temperature, the renormalisation factor may be calculated directly
\begin{equation}
Z\approx 1+\frac{\hbar}{N}\sum_{\bf k}
\frac{b_{n}J_{0n}^{o}({\bf k})}{(E_{0}-E_{n}-\hbar\omega_{{\bf k}o})^2}, 
\end{equation}
the sum rule $|A(t=0)|=1$ enables us to
 estimate the effective damping constant $\alpha$. At higher temperatures,
 when the phonon dispersion effects are strong we evaluate $Z$
 using the same condition ($|A(t=0)|=1$).
   
\section{Results and conclusions}

We have evaluated the interaction form factors
 $J_{nn^{'}}^{s}({\bf k})$
 taking the function of the non--interacting electron and hole system 
 as the first approximation for the exciton wave function 
 $\Phi_{n}({\bf r}_{e},{\bf r}_{h})=
 \Psi^{(e)}_{n_{e}m_{e}}({\bf r}_{e})
 \Psi^{(h)}_{00}({\bf r}_{e})$, where
\begin{eqnarray}
\Psi^{(i)}_{nm}({\bf r})=
\psi^{(i)}_{\bot nm}(r_{\bot},\varphi)\phi^{(i)}(z)
=\frac{N_{nm}^{(i)}}{(2\pi)^{1/2}}
\left(\frac{r_{\bot}^{2}}{l_{i}^{2}}\right)^{|m|/2}
{\rm e}^{-\frac{r_{\bot}^{2}}{2l_{i}^{2}}}
L_{n}^{|m|}\left(\frac{r_{\bot}^{2}}{l_{i}^{2}}\right)
{\rm e}^{{\rm i}m\varphi}
\frac{1}{\pi^{1/2}l_{iz}}
{\rm e}^{-\frac{z^{2}}{2l_{iz}^{2}}},
\nonumber\\
N_{nm}^{(i)}=\frac{1}{\pi^{1/2}l_{i}}\left(\frac{n!}{(n+|m|)!}\right).
\end{eqnarray}
Since $\omega_{0}^{i}\ll\omega_{z}^{i}$, we consider only the first level 
 for the quantisation in z--axis direction. A fact that an effective mass 
 of a hole is large in comparison to the electron effective mass, enables one 
 to neglect the 
 hole levels different than the lowest one (of $n_{h}=0$, $m_{h}=0$).
 Defining the functions
\begin{eqnarray}
I_{n^{'}n^{''}}^{(i)}({\bf k})=
\int_{0}^{\infty}\int_{0}^{2\pi}
\psi^{(i)*}_{n^{''}}(r_{\bot},\varphi)
{\rm e}^{{\rm i}k_{\bot}r_{\bot}\cos(\varphi)}
\psi^{(i)}_{n^{'}}(r_{\bot},\varphi)
{\rm d}r_{\bot}{\rm d}\varphi
\nonumber\\
\times
\int_{-\infty}^{\infty}
\phi^{(i)*}(z){\rm e}^{{\rm i}k_{z}z}
\phi^{(i)}(z){\rm d}z
=I_{\bot n^{'}n^{''}}^{(i)}(k_{\bot})
{\rm e}^{-2(k_{z}l_{iz}/2)^{2}},
\label{Iexp}
\end{eqnarray}
where the index $n$ replaces both $n_{e}$,$m_{e}$ quantum numbers,   
 we write the form factors in the form
\begin{eqnarray}
J_{nn^{'}}^{o}({\bf k})=
\frac{2\pi e^{2}\hbar\Omega}{\upsilon\tilde{\epsilon}k^{2}}
[I_{nn^{'}}^{(e)}({\bf k})-I_{nn^{'}}^{(h)}({\bf k})]^{2},
\nonumber\\
J_{nn^{'}}^{a}({\bf k})=
\frac{\hbar k}{2Mc_{a}}[\sigma_{e}I_{nn^{'}}^{(e)}({\bf k})-
\sigma_{h}I_{nn^{'}}^{(h)}({\bf k})]^{2}.
\end{eqnarray}

We perform the calculations of the energy shift including ten  
 lowest exciton energy levels, for the two confinement sizes
 $l_{e}=6$nm$\approx l_{h}$ and $l_{e}=3$nm$\approx l_{h}$.
 The following material parameters suitable to the InAs/GaAs QD have been used  
 $m^{*}_{e}=0.067m_{0}$, $m^{*}_{h}=0.38m_{0}$, $\epsilon_{0}=12.9$,
 $\epsilon_{\infty}=10.9$, $\hbar\Omega=36.4$meV, $c_{a}=4.8\cdot 10^{3}$m/s. 
 The integral form factors of the exciton--LO phonon interaction 
 $J_{0n}^{o}=(\hbar/N)\sum_{\bf k}J_{0n}^{o}({\bf k})$ 
 multiplied by the level degeneracy rank corresponding
 to the exciton energies $E_{n}$ are presented in the table.

The inclusion of many exciton levels influences the
 dressed exciton energy shift. It is responsible for its
 substantial increase (from 0.5meV found in one level approximation
 to 3.6meV when include ten levels for $l_{e}=3$nm,
 and from 0.35meV to 2.9meV for $l_{e}=6$nm). The
 other material constants $\sigma_{e}=6.7$eV, $\sigma_{e}=2.7$eV, 
 $\rho=5.36$g/cm$^{3}$ (Adachi 1985) are useful for estimations.
 
The spectral density and its inverse Fourier transform 
 calculated including many exciton levels and including realistic 
 asymmetry of the QD ($l_{ze}/l_{e}=1/3$) are plotted in Figs 1,2.
 Generally, the LA channel of dressing
 gives for the typical QDs the {\em picosecond scale of dressing}.
 Inclusion of the LO channel 
 does not modify significantly the overall LO and LA dressing kinetics
 in comparison to LA channel solely (Takagahara 1999).
 Thus the LA channel of dressing gives
 the limit for adiabatic creation of exciton--polaron in InAs/GaAs QD.
 The slope of the time courses 
 of spectral functions is related to damping of the dressed exciton
 at non zero temperature. The relaxation time is strongly influenced 
 by the size of QD. 

Let us emphasise that our results for time dependence of the response 
 functions (Fig. 1,2) coincide with the experimental results by Borri
 (2001).\\

\vspace*{2ex}

The work was supported by E.C.Project IST-1999-11311 (SQID)
 and by KBN Project PB 2 PO3B 055 18.
 
\newpage
\section*{References}
\noindent
Adachi S, J. Appl. Phys. {\bf 58}, 1 (1985)\\
Borri P, Langbein W, Woggon U, Sellin RL, Ouyang D,
Bimberg D, Phys. Rev. Lett. {\bf 87}, 157401 (2001)\\
Davydov AS, Pestryakov GM, Phys. Stat. Sol. B {\bf 49}, 505 (1972)\\
Davydov AS, "Solid State Theory" Nauka (1976) (in Russian)\\ 
Elliott RJ, Phys. Rev. {\bf 108}, 1384 (1957)\\
Engelsberg S, Schrieffer JR, Phys. Rev. {\bf 131}, 993 (1963)\\
Kikkawa JM, Awshalom DD, Phys. Rev. Lett. {\bf 80}, 4313 (1988)\\
Krummheuer B, Axt VM, Kuhn T, Phys. Rev. B {\bf 65}, 195313, (2002)\\
Langreth DC, Kadanoff LP, Phys. Rev. {\bf 133}, A1070 (1964),\\
Mahan GD, Phys. Rev. {\bf 142}, 366 (1966)\\
Mahan GD, "Many -- Particle Physics", Plenum (2000)\\
Martin PC, Schwinger J, Phys. Rev. {\bf 115}, 1342 (1959)\\ 
Moskalenko SA, Schmigluk MI, Chinik BI, Fiz. Tverd. Tela
{\bf 10}, 351 (1968) [Sov. Phys. -- Solid State]\\
Suna A, Phys. Rev. {\bf 135}, A111 (1964)\\
Takagahara T, Phys. Rev. B {\bf 60}, 2638 (1999)

\newpage

\noindent
Figure captions:\\
\noindent
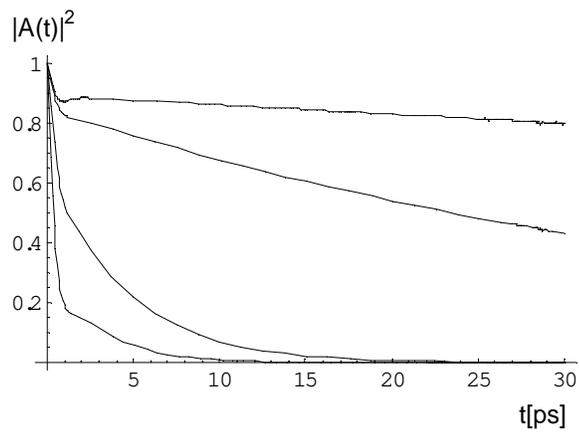
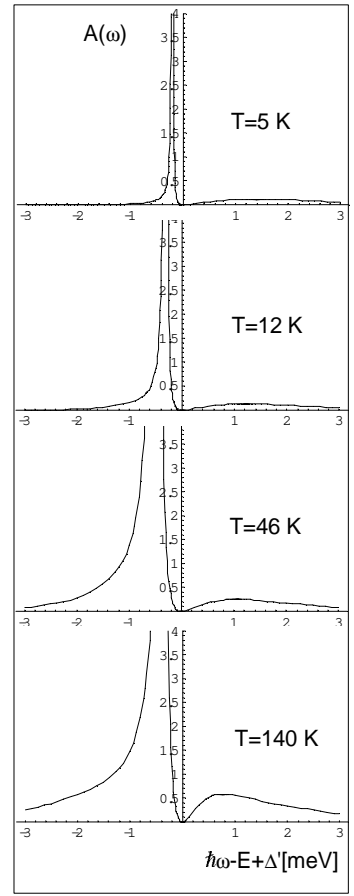
Fig. 1. Time course of the response function and spectral intensity 
for $l_{e}=3$nm, $l_{z}/l_{e}=1/3$\\
\noindent
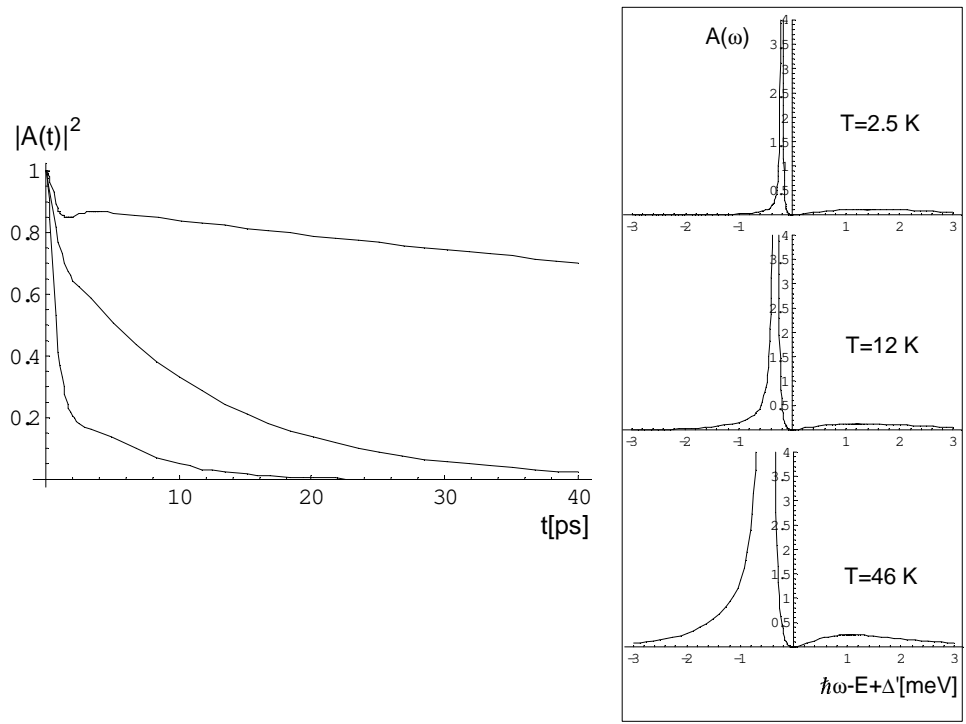
Fig. 2. Time course of the response function and spectral intensity 
for $l_{e}=6$nm, $l_{z}/l_{e}=1/3$\\

\noindent
Table caption:\\
\noindent
Exciton energy levels $E_{n}$ and corresponding integral form factors
$J_{0n}^{o}$ multiplied by n-th level degeneracy rank $b_{n}$.

\newpage

\vspace*{30ex}
\begin{center}
\begin{tabular}{|c|c||c|c|}
\hline
\multicolumn{2}{|c||}{$l_{e}=3$nm} & \multicolumn{2}{|c|}{$l_{e}=6$nm}\\
\hline
$E_{n}$[meV] & $b_{n}J_{0n}^{o}$[meV$^{2}$] &
$E_{n}$[meV] & $b_{n}J_{0n}^{o}$[meV$^{2}$] \\
\hline
88.9 & 13.1 & 7.87 & 12.5\\
113.3 & 2$\cdot$ 39.0 & 15.7 & 2$\cdot$18.8\\
134.0 & 36.0 & 23.3 & 21.5\\
135.8 & 2$\cdot$ 16.9 & 23.8 & 2$\cdot$9.9\\
152.0 & 2$\cdot$ 18.4 & 29.6 & 2$\cdot$10.4\\
155.6 & 2$\cdot$ 5.6 & 30.8 & 2$\cdot$3.5\\
167.9 & 18.2 & 34.8 & 7.4\\
169.4 & 2$\cdot$ 10.7 & 35.5 & 2$\cdot$4.7\\
183.8 & 2$\cdot$ 13.0 & 39.5 & 2$\cdot$5.4\\
199.0 & 11.5 & 44.2 & 4.0\\
\hline
\end{tabular}
\end{center}

\end{document}